\documentclass{emulateapj}

\shorttitle{Long Secondary Periods and Binarity in Red Giant Stars}
\shortauthors{Soszy{\'n}ski}

\begin{document}

\title{Long Secondary Periods and Binarity in Red Giant Stars}

\author{I. Soszy{\'n}ski}
\affil{Warsaw University Observatory, Al.~Ujazdowskie~4, 00-478~Warszawa, Poland}
\email{soszynsk@astrouw.edu.pl}

\begin{abstract}
Observational arguments supporting the binary explanation of the long
secondary periods (LSP) phenomenon in red giants are presented. 
Photometry of about 1200 semiregular variables with the LSP in the Large
Magellanic Cloud are analyzed using the MACHO and OGLE photometry. For
about 5\% of these objects additional ellipsoidal-like or eclipsing-like
modulation with the same periods as the LSP is detectable. These
double-humped variations are usually shifted in phase comparing to the LSP
light curves. I discuss the model of binary system with a red giant as the
primary component and a low-mass object as the secondary one. The mass lost
by the red giant through the wind follows the spiral pattern in the orbit
around the primary star and obscures it causing the LSP variations.
\end{abstract}

\vspace{0.3cm}
\section{Introduction}

Among numerous classes of variable stars only one type of large-amplitude
stellar variability remains completely unexplained. This is the long
secondary periods (LSP) observed in luminous red giant stars.  The LSP
variability with periods between 200 and 1500 days and with $V$-band
amplitudes up to 1~mag occurs in $\sim30$\% of Semiregular Variables (SRVs)
and OGLE Small Amplitude Red Giants (OSARGs). This phenomenon has been
known for decades \citep{pg54,hou63}, but an interest in the stars with the
LSP has been renewed since \citet{woo99} showed that these objects follow a
period--luminosity (PL) relation (sequence D).

In recent years our knowledge about the LSP phenomenon has significantly
increased, although its origin remains a mystery. \citet{hin02} and
\citet{woo04} studied spectral features of several Galactic stars with the
LSP and detected radial velocity variations with amplitudes of a few km/s
which agree with photometric long-period variations. \citet{sos04} used
Optical Gravitational Lensing Experiment (OGLE) data to select and analyze
close binary systems in the Large Magellanic Cloud (LMC) with a red giant
as one of the components. They noticed that the sequence D in the PL
diagram overlaps and is a direct continuation of the PL sequence of
ellipsoidal and eclipsing red giants\footnote{if the real orbital periods
of binary variables are considered, i.e. periods two times longer than
obtained automatically.} (sequence E), suggesting that the LSP phenomenon
is related to binarity. \citet{sos04} also showed that some evident
ellipsoidal red giants exhibit simultaneously OSARG-type variability, thus,
it was directly demonstrated that in some cases the binary explanation of
the LSP is true. Nevertheless, there is no doubt that the bulk of the LSP
variables are not typical ellipsoidal or eclipsing binaries.

\citet{sos05} discovered that the sequence D split into two ridges in the
period -- Wesenheit index ($W_I=I-1.55(V-I)$) plane, which corresponds to
the spectral division into oxygen-rich and carbon-rich AGB stars. The same
feature was noticed for Miras and SRVs (sequences C and C$'$), and the
sequence D contains relatively much smaller number of carbon-rich stars.

Recently \citet{der06} presented a period--luminosity--amplitude analysis
of variable red giants in the LMC. They examined amplitudes of ellipsoidal,
LSP and Mira-like variables using MACHO red ($R_M$) and blue ($B_M$)
photometry. The amplitude distribution for the LSP stars turned out to be
different than for ellipsoidal and pulsating variables, but blue-to-red
amplitude ratios of the LSP stars (typically 1.3) is more similar to this
quantity in pulsating variables ($\sim1.4$) than to ellipsoidal/eclipsing
binaries ($\sim1.1$). This last feature is used in present work for
separation of the LSP and ellipsoidal/eclipsing variability in the same
light curves.

Various hypotheses have been proposed to explain the origin of the LSP
variability: rotation of a spotted star, episodic dust ejections, a radial
and non-radial pulsation and binary companions including planets or brown
dwarfs. \citet{woo99} suggested that the sequence D stars are components of
semidetached binary systems. The matter lost by the AGB star forms a dusty
cloud around the companion, and regularly obscures the primary component
causing the LSP variability. \citet{hin02} and then \citet{ow03} mentioned
that the radial velocity measurements are consistent only with the binary
or pulsation explanations of the LSP. \citet{woo04} ruled out the binary
hypothesis, because a short ($\sim1000$ yr) timescale on which the
companion should merge with the red giant. They suggested that the most
likely explanation of the LSP are low degree g$^+$ pulsation modes trapped
in the outer radiative layers of the star.

The main goal of this paper is to find observational evidences for or
against the binary explanation of the LSP. Since the PL sequence populated
by the ellipsoidal and eclipsing variables overlaps with the LSP sequence
\citep{sos04}, it should be possible to find stars revealing simultaneously
both types of variability. If periods are the same, the LSP phenomenon must
be related to binarity. If not, the LSP is presumably caused by another
reason.

\vspace{0.5cm}
\section{Data Analysis}

Since the LSP are sometimes as long as 1500 days, presented analysis is
based on observational data originated in two sources: MACHO and OGLE
surveys. Merged light curves from both projects covered 15 years of
observations: from 1992 to 2006. The sequence D stars selected by
\citet{sos04} in the OGLE database were cross-identified with objects
collected in the MACHO
archive\footnote{http://wwwmacho.mcmaster.ca/Data/MachoData.html}. I found
about 1200 counterparts of the previously selected variables. Then,
$R_M$-band MACHO observations have been merged with the OGLE points by
scaling amplitudes and shifting zero points of the photometry. The
parameters of this transformation have been found by the least square
fitting to the measurements obtained between 1997 and 2000, i.e., when both
projects observed the LMC fields at the same time.

\vspace{0.3cm}
\subsection{Searching for orbital periods different than the LSP}

Searching for ellipsoidal or eclipsing variability with different periods
than the LSP was a relatively easy procedure. For each object a third
order Fourier series was fitted to the LSP light curve and subtracted from
the points. Then, the period search was performed for the residual
data. This procedure was repeated until four additional periods per star
have been found.

\begin{figure}
\includegraphics[width=8.7cm]{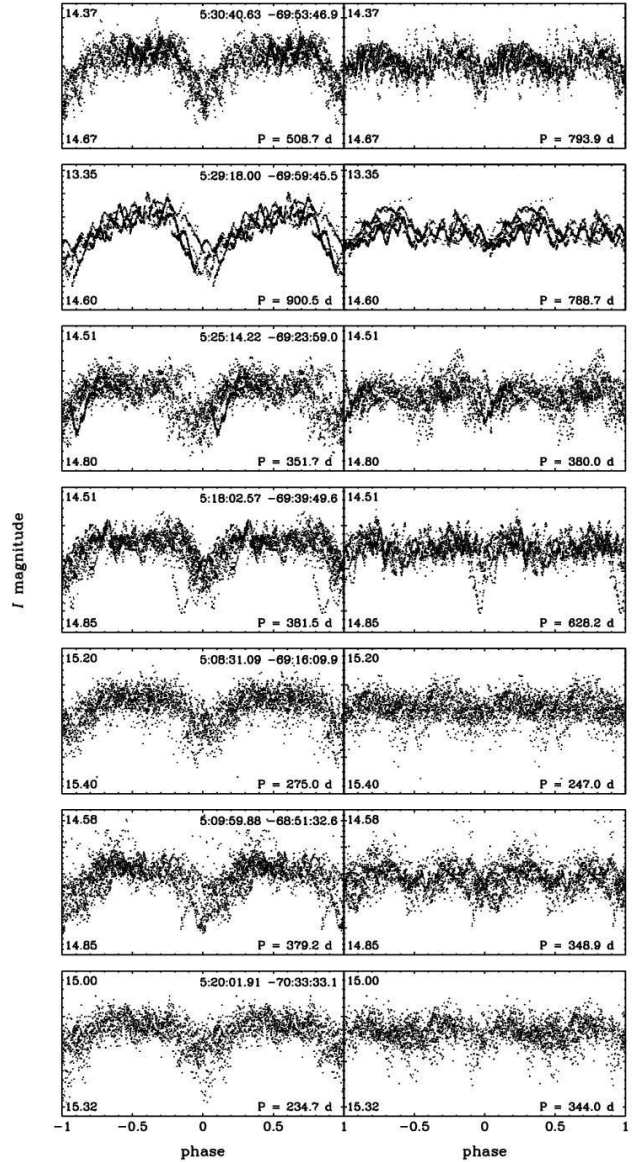}
\vspace{-0.5cm}
\caption{Light curves of LSP stars with secondary periods in the binarity
  range. Left column shows light curves folded with the LSP, and right
  column presents these data after subtracting the LSP variability and
  folded with different periods.}
\end{figure}

From these results I selected and visually inspected light curves with
periods lying between sequences C and D, i.e., where one can expect
ellipsoidal or eclipsing variables (automatic procedure gives a half of
orbital periods). The best seven light curves of that type are plotted in
Fig.~1. Phased LSP light curves are shown in the left column, the right
column presents the light curves of these objects after subtracting the LSP
variations and phased with periods two times longer than obtained
automatically.

As one can see none of these secondary modulations is distinct,
indisputable eclipsing or ellipsoidal light curve. The most likely
explanation of these modulations are variations of phases or amplitudes of
the LSP variability which produce such ``artificial'' variability in
residual light curves. A very similar behavior occurred in eclipsing binary
described in Section~3.

\vspace{0.3cm}
\subsection{Searching for orbital periods the same as the LSP}

If the LSP is related to binarity, one can expect that a number of objects
shows ellipsoidal or eclipsing variability of the same period as the LSP.
Unfortunately, detecting such a modulation in the LSP light curves is not
an easy task, because (i) the ellipsoidal light curves have usually much
smaller amplitudes than LSP, (ii) shorter semiregular variability is
superimposed, (iii) the LSP light curves often change amplitudes, phases
and periods. A careful investigation of the LSP light curves reveals that
some of them show shallow secondary minima, what may be a sign of
ellipsoidal variations superimposed on the LSP. However, since the LSP
light curves appear in several variants, it is possible that such behavior
is not related to binarity.

To separate LSP and possible ellipsoidal/eclipsing variations I used a
feature noticed by \citet{der06}. They studied MACHO $B_M$ and $R_M$
photometry of long period variables in the LMC and found that blue-to-red
amplitude ratios are different for ellipsoidal and LSP variables. In the
ellipsoidal red giants, where the variability is dominated by the geometric
changes, amplitudes in different filters are very similar. The median value
of $A(B_M)/A(R_M)$ is equal to 1.1. For the LSP stars $A(B_M)/A(R_M)$ is
more similar to pulsating variables and equal on average to 1.3. It means
that if the amplitudes of the $R_M$-band light curves are scaled by a
factor 1.3 and the $B_M$ observations are subtracted, the LSP variations
will be canceled, but not the possible ellipsoidal/eclipsing modulation.

\begin{figure*}
\hspace{1cm}
\includegraphics[width=16cm]{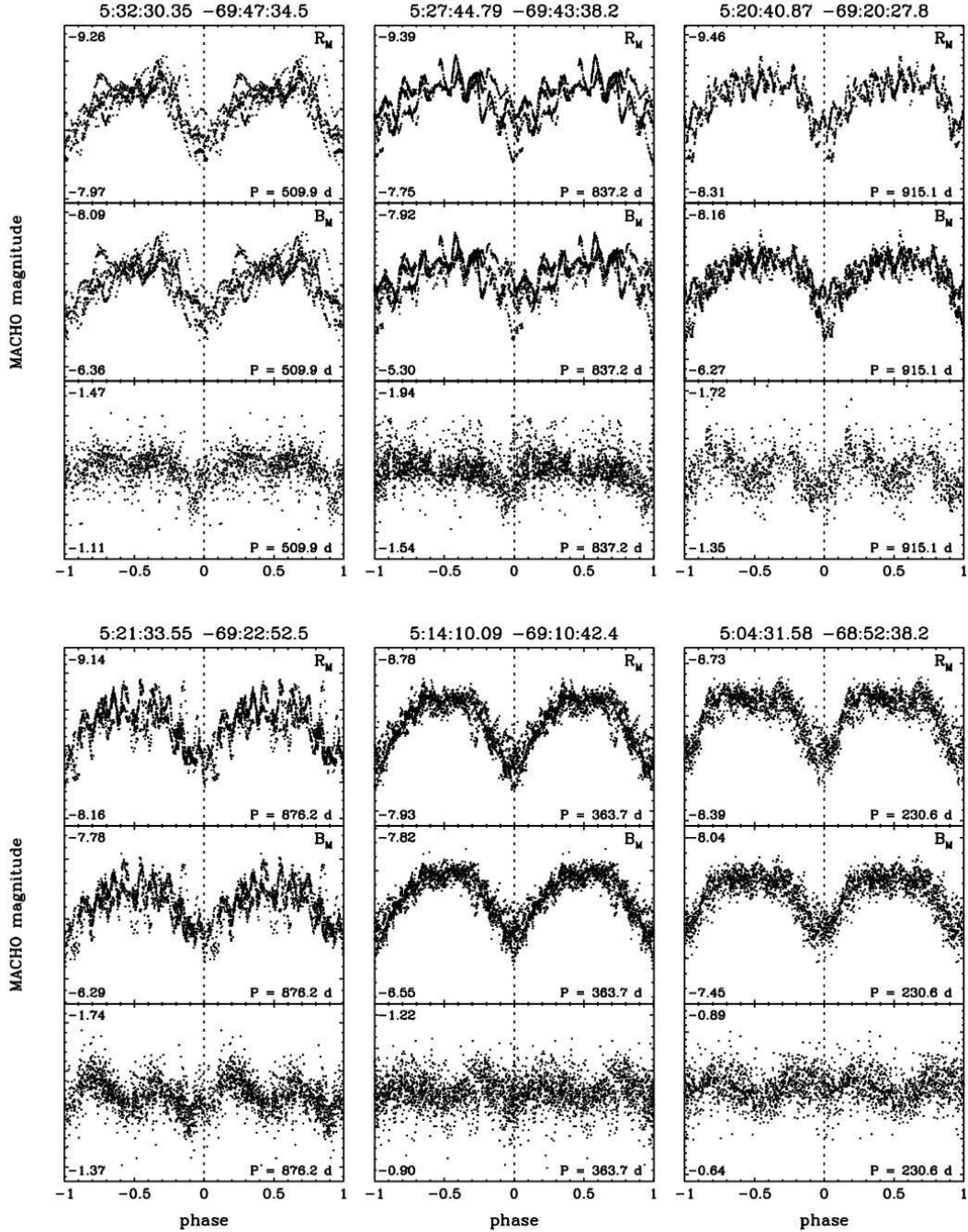}
\vspace{-2.5cm}
\caption{Six LSP star exhibiting additional double-humped
  variations. Upper diagram in each panel shows MACHO $R_M$-band
  photometry, middle diagram contains $B_M$-band data, lower diagram
  presents the residual data after scaling $R_M$ and subtracting $B_M$
  light curves. The dashed lines show the phases of minimum brightness
  during the LSP variations. Note that each light curve of given star is
  folded with the same period.}
\end{figure*}

The procedure was as follows. The $B_M$ and $R_M$-band amplitudes were
determined by fitting the spline function to the folded light curves. Then,
the $R_M$ magnitudes were converted into flux and linearly scaled to obtain
the same amplitude (in magnitudes) as in the $B_M$ bandpass. The flux was
again converted into magnitudes, and $B_M$ measurements were subtracted
point by point from scaled $R_M$ data. The final step of the procedure was
fitting a sinusoid of a period 1 year, and subtracting the function from
the residual data. This way I removed a differential refraction effect
distinctly visible in MACHO data.

For the vast majority of light curves the LSP and pulsation (semiregular)
variability subtracted very well. No significant long-period variability
was detectable in the residual points. However, for about 5\% of stars I
found clear periods equal to half the LSP, i.e. I noticed double-humped
curves with periods the same as of the LSP. All these residual data are
available from the electronic edition of the Astrophysical Journal.
The most prominent curves of this type are presented in Fig.~2. In these
cases the residual data seem to arrange in eclipsing-like or
ellipsoidal-like light curves.

Fig.~3 shows period--$W_I$ diagram (where $W_I=I-1.55(V-I)$ is the
reddening-free Wesenheit index) for stars from sequences D and E. Full
circles show the position of the LSP stars with the double-humped
curves. Note that these objects appear along the complete length of
sequence~D, although there is an overdensity for shorter-period stars,
i.e. where sequences D and E overlap. Note also that double-humped
objects populate both, O-rich and C-rich, sequences of the LSP stars
\citep{sos05}.

\begin{figure}
\vspace{-0.6cm}
\hspace{-2.0cm}
\includegraphics[width=12cm]{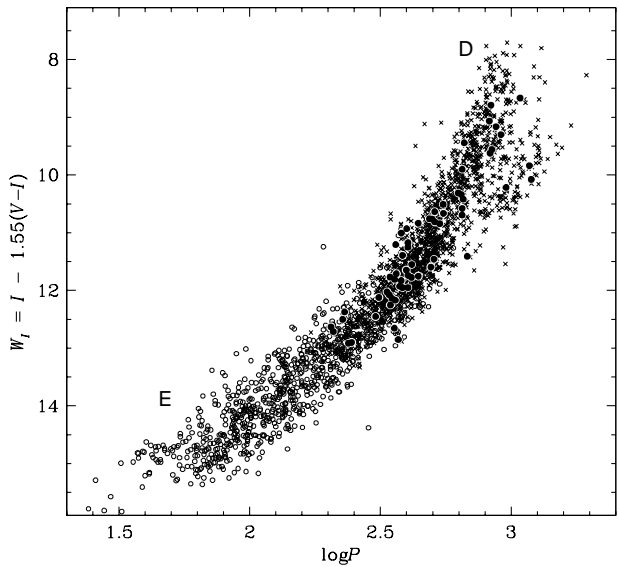}
\vspace{-8cm}
\caption{Period--$W_I$ diagram for the LSP stars and ellipsoidal red giants
  in the LMC. Crosses show the LSP stars (sequence D), empty circles
  indicate ellipsoidal red giants (sequence E), filled circles mark the LSP
  stars with double-humped curves visible in residual data.}
\end{figure}

It is worth mentioning that the described procedure does not demand any
assumptions concerning periods or light curve shapes. It is a simple
transformation, the same for each observing point. The only parameter of
this transformation is a blue-to-red amplitude ratio of the LSP
variability, but I checked that the double-humped residual light curves are
visible for relatively wide range of this parameter, so it is not very
important to measure amplitudes very accurately.

\vspace{0.3cm}
\section{Discussion}

I have shown that about 5\% of sequence D objects have ellipsoidal or
eclipsing-like modulation with periods the same as the LSP. In the sample
of 1200 sequence D stars I did not find any distinct ellipsoidal or
eclipsing binary with period different than the LSP. Although various
scenarios are still possible, it is justified to re-invoke the binary
hypothesis proposed by \citet{woo99}. In this explanation the AGB star in a
close binary system losses mass to the secondary component. The matter
forms a dusty cloud around and behind the companion and regularly obscures
the primary star.

An argument for this hypothesis is a phase lag between
ellipsoidal/eclipsing and the LSP variations clearly visible in
Fig.~2. Only one object of six -- the star with the shortest period -- does
not exhibit such behavior. I checked that it might be a rule for the
shortest period sequence D variables. For longer periods the minima of the
LSP brightness variation occur about 0.05--0.10 of a cycle after the minima
of the ellipsoidal/eclipsing light curves. Consistent results were
presented by \citet{woo01}, who discovered a phase lag of $\sim$1/8 between
LSP light curves and radial velocity curves. Such phase offsets between
ellipsoidal/eclipsing and LSP minima agree very well with hydrodynamical
simulations of a wind driven accretion flow in binary systems
\citep{tj93,mm98,nag04}. These models predict that a matter lost by a red
giant in a binary system follows the spiral pattern with maximum density
located behind the secondary component.

An exhaustive discussion about possible origins of the LSP was done by
\citet{woo04}. They argued that the merger timescale for red giants and its
companion in close binary systems is of the order of $10^3$ years, while
the lifetime in the AGB phase is two orders of magnitude longer. Thus, 30\%
of SRVs showing the LSP is highly inconsistent with the binary scenario.

However, the mass transfer in a binary system may be due to the Roche-lobe
overflow \citep[e.g.][]{pac71}, or through the stellar wind
\citep[e.g.][]{abm87,tj93}. The former process tends to circularize the
orbits and to synchronize the spin of the stars with the orbital rotation,
which results in shrinking the orbits and finally in merging the
components. The latter phenomenon increases the eccentricity of the orbits
and, if the bulk of the mass lost by the red giants escapes from the system,
the distance between components may even increase. Thus, the main argument
of \citet{woo04} against the binary explanation may be not valid, if we
assume that the red giants in the binary systems do not fully fill their
Roche lobe, and the bulk of mass transfer is driven by the stellar
wind. The non-sinusoidal radial velocity curves observed by \citet{hin02}
and \citet{woo04}, which can be explained by the eccentricity of the
orbits, are in good agreement with this scenario.

\begin{figure*}
\includegraphics[width=17cm]{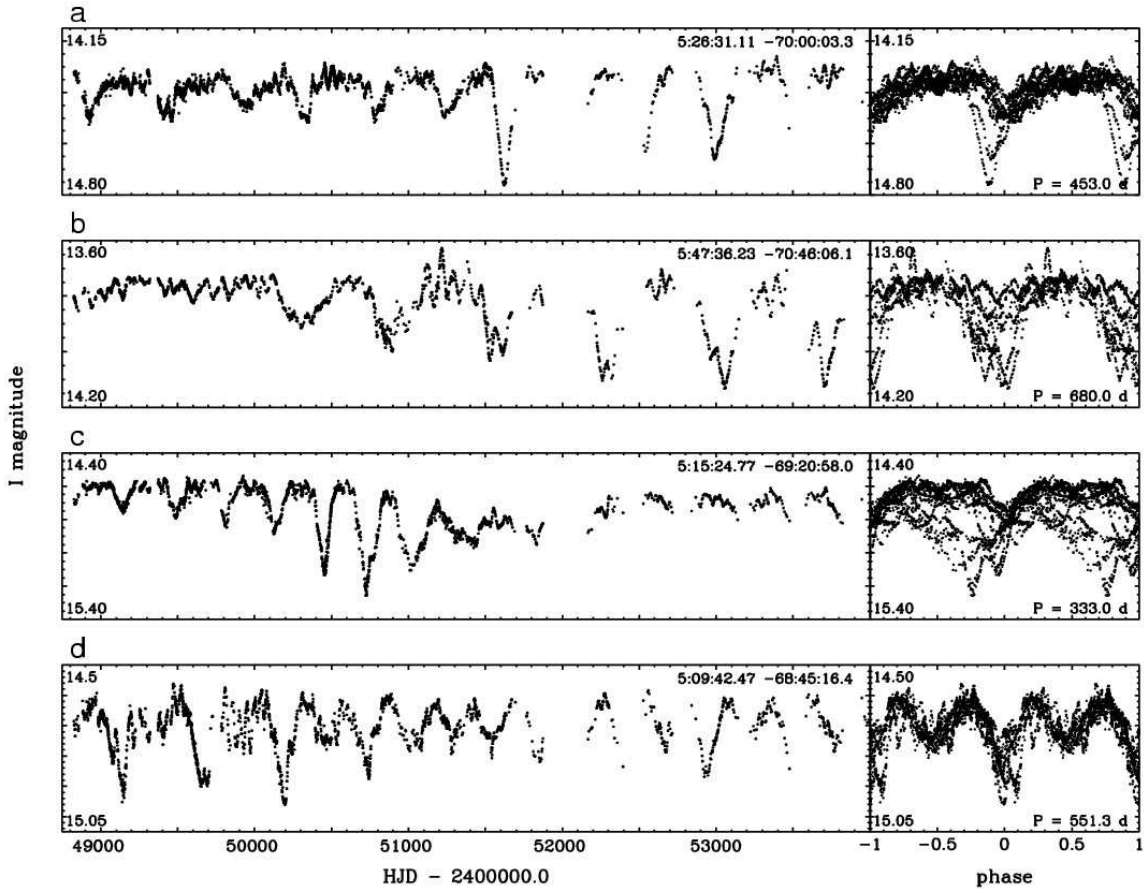}
\vspace{-0.5cm}
\caption{Three upper panels show LSP stars with variable amplitudes. In the
  lower panel the probable eclipsing binary is presented. Left and right
  diagrams contain unfolded and folded light curves, respectively.}
\end{figure*}

Presented hypothesis nevertheless requires that at least 30\% of AGB stars
exist in close binary systems. Moreover, while the studied LSP variables
have velocity amplitudes of only a few km/s \citep{hin02,woo04}, the
confirmed binary systems (sequence~E stars) appear to have velocity
amplitudes about ten times larger \citep{awc06}. The radial velocity
measurements suggest that the second component in the LSP stars may be a
brown dwarf. \citet{ret05} proposed that the Jupiter-like planets may
accrete the matter from its host star and increase mass into the brown
dwarfs range. This hypothesis would explain such large number of the LSP
cases among AGB stars, if we assume that planets with a separation of 1--5
AU are common. To test the binary hypothesis it would be interesting to
measure the radial velocity changes for the LSP stars with the
double-humped variations. If these stars are ellipsoidal or eclipsing
variables indeed, their velocity amplitudes should be similar to these
observed in sequence E stars, i.e. significantly larger than for the
remaining LSP stars.

The brown dwarf scenario can also explain why only 5\% of our sample show
the double-humped variability. For low-mass secondary components the
ellipsoidal and eclipsing variations have too small amplitudes to be
detected in our procedure. One should remember that the residual data
obtained by scaling $R_M$ and subtracting $B_M$ magnitudes can show
variability with amplitudes of about 0.2 of the original ellipsoidal
amplitudes ($R_M$ were scaled by a factor of 1.3 and $A(B_M)/A(R_M)$ is on
average 1.1 for ellipsoidal variables). Moreover, presented model does not
assume that the red giant fills entirely the Roche lobe, because the bulk
of the mass flow is through a stellar wind, so the separation between the
components can be too large to cause significant ellipsoidal variability.
Note also that observed number of possible ellipsoidal and eclipsing
variables among the LSP variables is in agreement with relative number of
sequence~E stars. The LSP modulation is observed for about 30\% of the AGB
stars, so 5\% of these objects gives about 1-2\% of the whole
population. Exactly the same relative number of ellipsoidal or eclipsing
variables is observed for fainter red giants \citep{sos04}, so one should
not expect to detect many more such objects among brighter stars.

Of course, it cannot be absolutely excluded that the double-humped
variations have different explanation than the binarity. However, the
non-binarity model of the LSP phenomenon has to explain why produced
residual light curves have ellipsoidal or eclipsing-like shapes, why it
appears only in a few percent of the sequence D stars, and why there is
phase lag between LSP and double-humped variations. All these facts can be
explained by the binary scenario.

Additional clues on the origin of the LSP can be given by an analysis of
typical and peculiar cases of the light curves. A common feature of the LSP
variations is the modulation of the amplitudes and phases. A few such light
curves are shown in in Fig.~4. It seems that the phase shifts are
correlated with the depth of minima -- the deeper minimum occurs earlier
than the shallower one (see the folded light curve in Fig.~4a). Sometimes
the LSP variations completely disappear or appear in red giants which did
not show such modulation before (Fig.~4b). The conclusion that can be drawn
from these cases is that the regular (with no LSP) apparent luminosity of
the red giants is the same as in the maximum apparent luminosity of the LSP
variations, i.e. the LSP phenomenon decreases the total luminosity of the
star. This fact must be taken into consideration by any theory of the LSP
variability.

The hypothesis of a binary system with a mass-losing AGB star seems to agree
with these results, because the obscuration by a cloud of matter reduces
the total luminosity of the star. The amplitude and phase changes are
likely connected with the variable mass loss rate. An interesting LSP light
curve is presented in Fig.~4c. Starting from the beginning of the
observations the depth and width of the minima were increasing. After
a few cycles this process affected the maximum of the light curve, and the
total apparent optical luminosity of the star dropped down. Then, the
object returned, more or less, to the previous stage. This behavior can be
interpreted as a sudden rise of a mass loss rate, which caused the whole
system to be hidden in the cloud of matter.

In Fig.~4d I show a probable eclipsing binary system with the AGB star as
one of the components. This object presents striking similarities to the
LSP stars. First, it is located in the sequence D in the PL
diagram. Second, the light curve changes the amplitudes and phases of
variations. Third, after subtracting a function fitted to the primary
(eclipsing) variability I obtained secondary long-period variations caused
probably by modulation of the primary period. The same behavior I observed
for the LSP stars presented in Fig.~1.

\vspace{0.3cm}
\section{Summary and Conclusions}

In this paper I show that careful analysis of available data may shed new
light on the nature of the last unexplained type of stellar
variability. Arguments in favor of the binary explanation of the LSP
phenomenon are as follows:
\begin{enumerate}
\item There are no reliable examples of the LSP stars with ellipsoidal or
  eclipsing variations with different periods.
\item At least 5\% of the sequence D stars exhibit ellipsoidal or
  eclipsing-like variability with the same period as the LSP.
\item Phase lag between ellipsoidal and LSP variations agrees well with
  models of wind accretion in binary systems.
\item The LSP light curves with variable amplitudes can be explained by a
  variable mass-loss rate in binary systems.
\end{enumerate}

The binary scenario can explain many features of the LSP variables. The PL
relation (which is a direct continuation of ``binary'' sequence E) is a
projection of a radius--luminosity relation for red giants. Positive
correlation between amplitudes of the LSP variability and mean luminosity
\citep{sos04,der06} may be caused by the increasing mass loss rate with
the luminosity. The dimming during the LSP minima are consistent with
the obscuration by a dusty cloud of matter orbiting the red giant. Finally,
the radial velocity variations are in agreement with eccentric motion of
the low-mass companion. The long-term project of radial velocity
measurements of selected sequence D stars have been recently finished
(P.~Wood, private communication). I expect that these data will
definitively solve the LSP problem.

\acknowledgments
{ I am deeply grateful to Professors W.~D.~Dziembowski, B.~Paczy{\'n}ski,
  A.~Udalski, and Dr.~Z.~Ko{\l}aczkowski for the careful and critical
  reading of the manuscript and many useful discussions. I thank the
  anonymous referee for helpful comments. The paper was supported by the
  Foundation for Polish Science through the ``Homing Programme''.

  This paper utilizes public domain data obtained by the MACHO Project,
  jointly funded by the US Department of Energy through the University of
  California, Lawrence Livermore National Laboratory under contract
  No. W-7405-Eng-48, by the National Science Foundation through the Center
  for Particle Astrophysics of the University of California under
  cooperative agreement AST-8809616, and by the Mount Stromlo and Siding
  Spring Observatory, part of the Australian National University.

  This publication is partly based on the OGLE observations obtained with
  the Warsaw Telecope at the Las Campanas Observatory, Chile, operated by
  the Carnegie Institution of Washington.}


\begin{thebibliography}{}
\bibitem[Adams et al.(2006)]{awc06} Adams, E., Wood, P.~R., \& Cioni, M.-R.\ 2006, Memorie della Societa Astronomica Italiana, 77, 537 
\bibitem[Anzer et al.(1987)]{abm87} Anzer, U., Boerner, G., \& Monaghan, J.~J.\ 1987, \aap, 176, 235
\bibitem[Derekas et al.(2006)]{der06} Derekas, A., Kiss, L.~L., Bedding, T.~R., Kjeldsen, H., Lah, P., \& Szab{\'o}, G.~M.\ 2006, \apjl, 650, L55
\bibitem[Hinkle et al.(2002)]{hin02} Hinkle, K.~H., Lebzelter, T., Joyce, R.~R., \& Fekel, F.~C.\ 2002, \aj, 123, 1002
\bibitem[Houk(1963)]{hou63} Houk, N.\ 1963, \aj, 68, 253
\bibitem[Mastrodemos \& Morris(1998)]{mm98} Mastrodemos, N., \& Morris, M.\ 1998, \apj, 497, 303
\bibitem[Nagae et al.(2004)]{nag04} Nagae, T., Oka, K., Matsuda, T., Fujiwara, H., Hachisu, I., \& Boffin, H.~M.~J.\ 2004, \aap, 419, 335
\bibitem[Olivier \& Wood(2003)]{ow03} Olivier, E.~A., \& Wood, P.~R.\ 2003, \apj, 584, 1035
\bibitem[Paczy{\'n}ski(1971)]{pac71} Paczy{\'n}ski, B.\ 1971, \araa, 9, 183 
\bibitem[Payne-Gaposhkin(1954)]{pg54} Payne-Gaposhkin, C.\ 1954, Harvard Annals, 113, No. 4
\bibitem[Retter(2005)]{ret05} Retter, A.\ 2005, Bulletin of the American Astronomical Society, 37, 1487 
\bibitem[Soszy{\'n}ski et al.(2004)]{sos04} Soszy{\'n}ski, I., Udalski, A.,
 Kubiak, M., Szyma{\'n}ski, M.~K., Pietrzy{\'n}ski, G., {\.Z}ebru{\'n}, K.,
 Szewczyk, O., Wyrzykowski, {\L}., \& Dziembowski, W.~A.\ 2004, Acta
 Astron., 54, 347
\bibitem[Soszy{\'n}ski et al.(2005)]{sos05} Soszy{\'n}ski, I., Udalski, A., Kubiak, M., Szyma{\'n}ski, M.~K., Pietrzy{\'n}ski, G., {\.Z}ebru{\'n}, K., Szewczyk, O., Wyrzykowski, {\L}., \& Ulaczyk, K.\ 2005, Acta Astron., 55, 331
\bibitem[Theuns \& Jorissen(1993)]{tj93} Theuns, T., \& Jorissen, A.\ 1993, \mnras, 265, 946
\bibitem[Wood et al.(1999)]{woo99} Wood, P.~R., et al. 1999, in IAU Symp.~191: Asymptotic Giant Branch Stars, eds. T.~Le~Bertre, A.~L{\'e}bre, and C.~Waelkens (San Francisco: ASP), 151
\bibitem[Wood et al.(2001)]{woo01} Wood, P.~R., Axelrod, T.~S., \& Welch, D.~L.\ 2001, ASP Conf.~Ser.~229: Evolution of Binary and Multiple Star Systems, eds. Ph.~Podsiadlowski, S.~Rappaport, A.~R.~King, F.~D'Antona, and L.~Burder, 233
\bibitem[Wood et al.(2004)]{woo04} Wood, P.~R., Olivier, E.~A., \& Kawaler, S.~D.\ 2004, \apj, 604, 800

\end{thebibliography}
\end{document}